\documentstyle[amssymb,preprint,aps]{revtex}

\begin{document}
\draft

\noindent {\bf ELECTRONIC\ TRANSPORT IN\ La-Ca MANGANITES}

\vspace{0.5in}

\noindent Marcelo Jaime

\noindent Los Alamos National Laboratory

\noindent Los Alamos, NM 87545

\vspace{0.25in}

\noindent Myron B. Salamon\hspace{1in}

\noindent Department of Physics and Materials Research Laboratory.

\noindent University of Illinois at Urbana-Champaign.

\noindent Urbana, IL 61801

\vspace{0.25in}

\noindent Plenum Press, to be published.\hspace{1in}

\vspace{1.5cm}

\begin{center}
INTRODUCTION{\bf \ }
\end{center}

\tightenlines

\medskip

So called colossal magnetoresistance (CMR) manganites of composition A$%
_{1-x}^{3+}$B$_x^{2+}$MnO$_3$ (A: La, Nd; B: Ca, Sr, Pb) have been subject
of intense research during the last years, especially after their
re-discovery\cite{chahara}, in part due to their potential use as device
applications in the magnetic storage industry, and in part because of the
complexity of the mechanisms responsible for unusual, interdependent
electric, magnetic and structural properties. The materials and some of
their transport properties, however, have been known since the early
experiments by Jonker and van Santen\cite{jonker}, and Volger\cite{volger}
in the early fifties. Somewhat later, resistivity, magnetoresistance,
specific heat and magnetization were studied as functions of temperature in
La$_{1-x}$Sr$_x$MnO$_3$ polycrystalline samples. The transport data show
clearly most of the relevant physics that we know today, including a
surprising disagreement between the thermoelectric and Hall effects
regarding the sign of the charge carriers, accompanied by a quite small Hall
mobility. Volger interpreted the simultaneous metal-insulator and
ferromagnetic phase transitions observed in his samples in terms of the
double exchange theory (DE) for metallic ferromagnets first suggested by
Zener\cite{zener} and, perhaps worried about specimen quality issues,
intergrain-barrier effects, neglected the clues that his data provided in
support of some lattice involvement. This trend persisted over the ensuing
years, inspiring both theoretical models \cite
{anderson,mannari,kasuya,degennes,kubo,mazaferro,furu1,varma,allub} and the
interpretation of experimental results \cite{wollan,kusters,clausen,vonH}.

We revisit the transport properties of CMR in detail below, since they have
been confirmed in a number of CMR-exhibiting materials, and present new data
considered the hallmark of an unusually high electron-phonon coupling
responsible for charge localization and small polaron transport. The
discussion is organized in four different sections. In the first place we
briefly review materials and phase diagrams, as well as general properties,
with remarks on differences between intrinsic and extrinsic properties.
Second, the high temperature properties are analyzed, emphasizing those
which more clearly show the role of the coupling to the lattice. Next, the
very low temperature limit is considered, were the double exchange physics
is believed to rule. Finally, in the fourth section, an statistical model
for the phase transitions of manganites is presented, in an attempt to
bridge the far less well understood region of intermediate temperatures with
a reasonable extrapolation of what we know and understand in the extremes.
We will see that transport experiments give us again an important insight
into the dominant physics, that of localized charge carriers that reinforce
the tendency toward magnetic order when they gain mobility through the
ferromagnetic transition. Most of the discussion is restricted to the doping
region $x$ \symbol{126}1/3, where the low temperature metallic properties
are optimized.

\vspace{1cm}

\begin{center}
GENERAL PROPERTIES
\end{center}

\vspace{0.5cm}

LaMnO$_3$ is a cubic antiferromagnetic (AFM) semiconductor perovskite, with
Neel temperature $T_N\approx 140$ K, where magnetic moments at Mn sites are
ferromagnetically (FM) coupled in planes that alternate spin orientations in
what is known as A-phase\cite{wollan}, as displayed in Fig. 1a. In this
structure Mn$^{3+}$, surrounded by six oxygen atoms, is a Jahn-Teller (JT)
atom. The $d$-shell electronic energy levels $t_{2g}$ (triplet) and $e_g$
(doublet), in consequence, split under a distortion of the octahedrally
coordinated Mn-O bonds. The JT splitting reduces the electronic energy as
schematically shown in Fig. 1c. Three strongly coupled and localized ($%
t_{2g} $) electrons, occupy the bottom-most levels and form the core spin $%
S=3/2$. The fourth electron, occupying the first $e_g$ level, is coupled to
the core spin through the intra-atomic Hund's coupling constant $J>0$,
estimated on the scale of \symbol{126}1 eV.

This apparently simple system is by itself a materials-science challenge. In
first place pure and ideal LaMnO$_3$ crystals are in principle
semiconductors, while real laboratory samples are often good conductors that
show a variety of magnetic structure and ordering. This behavior originates
in unavoidable cationic (La$^{3+}$) vacancies that naturally occur during
sample preparation\cite{raveau,cnrrao,mahendiran} and introduce charge
carriers, magnetic frustration and lattice stress in the system. Cationic
vacancies in particular and deviations from ideal stoichiometry in general
are the most common causes for discrepancy among experimental results
reported in the literature. Associated with cationic vacancies it is not
unusual to observe also oxygen deficiency.

Band calculations using density-functional methods (LDA) predicted a
metallic ground state for an hypothetical cubic/undistorted version LaMnO$_3$%
, a finding that is at odds with the experimental results. Satpathy {\it et
al}.\cite{satpathy} have investigated and identified the physical reason for
this behavior by introducing different distortions into the oxygen
octahedrons. They have studied three different distortions, {\it i.e}. the
breathing mode (Q1), the basal-plane distortion mode (Q2) displayed in Fig.
1b, and a stretching mode (Q3) in addition to a small rotation of the
octahedron. They claim that ``{\em for LaMnO}$_3${\em \ a Jahn-Teller
distortion of the Q2 type, with the basal-plane oxygen atoms displaced by at
least the amount }$\approx ${\em \ 0.1 \AA\ from their ideal positions, is
necessary for an insulating band structure within the LDA and that the Q1 or
the Q3 distortions are not effective in opening up the gap.}'' They also
argue that Jahn-Teller (JT) like distortions favor antiferromagnetic rather
than ferromagnetic order. The implications of these findings are quite
relevant. Even though they only discuss static distortions, its clear that
the electronic band structure is extremely sensitive to particular phonon
modes and that a large enough distortion of the Mn-O bonds can drive the
system through a metal-insulator phase transition.

Another surprising, and has turned out to be, important characteristic of
LaMnO$_3$ is that hole doping by means of chemical substitution of La$^{3+}$
with Ca$^{2+}$ or Sr$^{2+}$, while increasing the electrical conductivity,
does not always produce metallic samples. Again, a rich spectrum of magnetic
ordering and/or charge localization has been found experimentally. In
particular, as much as 30\% content of B$^{2+}$ is required to observe a
metallic behavior at room temperature, as it can be seen in the phase
diagram in Fig 2.

When a divalent atom replaces La in the structure, electrical neutrality is
granted by the mixed valence nature of Mn. Indeed, Mn$^{3+}$ gives up one
electron per dopant atom in order to keep oxygen happy, resulting in $x$ Mn$%
^{4+}$ atoms per formula. As a result, a random elastic strain field is
introduced in the lattice. Because $e_g$ levels are empty in Mn$^{4+}$ ions,
they cannot profit from the JT effect; there is no energy gain obtained in
the $e_g$ level and the distortion is no longer favored. In consequence the
lattice long-range periodic distortion is now frustrated by a non-JT atom,
as schematically displayed in Fig. 3. Holes are likely to stay localized at
those Mn$^{4+}$ sites, since some elastic energy must be paid to move them
into a Mn$^{3+}$ site, and eventually they reach some kind of periodic
distribution, generating charge ordering in the system as the density of
holes increases. This would likely be the scenario in the case of spin-less
charge carriers. In the case of the manganites, however, the spin-1/2 holes
move in a spin-2 environment resulting in a remarkable enrichment of the
physics and phenomenology. Indeed, for a critical concentration of holes
experimentally found to be close to $x=1/3$, and temperatures low enough to
keep the spin fluctuations small, holes improve their jump probability,
reducing their kinetic energy, while moving between Mn sites with core $%
t_{2g}$ moments that point toward the same direction. This is direct
consequence of a very strong Hund's rule at Mn atoms, and is the essence of
the double exchange mechanism. When these clusters of magnetically aligned
Mn are large enough to overlap the same holes that benefited form local
magnetic order, delocalize acting as the driving force for global
ferromagnetism and a phase transition into a FM metallic state.

A large electron-phonon coupling is evidenced by an overwhelming amount of
experimental data. Outstanding unambiguous evidence of coupling to
Jahn-Teller lattice distortions include large pressure effects \cite
{khazeni,neumeier,hwang1}, magnetostriction effects associated with the
metal-insulator transition \cite{ibarra,rada1,dai,teresa}, a discrepancy
between the chemical potential estimation by means of thermoelectric effects
and thermally activated behavior of the electrical conductivity \cite
{tanaka,mj1,crespi,mj2,zhou,palstra,hundley}, the sign anomaly of the Hall
effect, and the Arrhenius behavior of the drift and Hall mobilities\cite{mj3}
. Further, optical properties\cite{kaplan}, charge ordering observed in the
low doping limit \cite{yamada}, local atomic structure studies\cite{billinge}%
, neutron diffraction studies \cite{rada2,louca1}, isotope effect \cite
{zhao1,zhao2,Isaac}, X-ray absorption fine-structure measurements that
indicate delocalization of charge carriers at the Curie temperature ($T_C$)
as well as coupling between distortions, charge distribution and magnetism%
\cite{booth1,booth2}, electron paramagnetic resonance\cite{shenge}, thermal
transport \cite{cohn}, and muon spin relaxation ($\mu $SR)\cite{heffner} add
to the list.

The transport properties can be discussed qualitatively with the help of a
resistivity vs. temperature curve. We will use the simple diagram in Fig. 4
to illustrate them. The temperature range is divided in three regions, {\it %
e.g}. much lower than $T_C$ (I), much higher than $T_C$ (II), and around to $%
T_C$ (III). The most important energy scale in this diagram is obviously $T_C
$, which is determined by $i$) bandwidth, $ii$) band filling, $iii$) local
disorder, and $iv)$ effective electron-phonon coupling. The bandwidth is
fundamentally fixed once the structure of the system is fixed, {\it e.g}.
atomic radii determine the structure and an average Mn-O-Mn bond angle for a
particular composition (A, B, $x$). A clever way to characterize the
structure is to use the so called ''tolerance factor'' $tf=\overline{\text{%
A-O}}/\sqrt{2\overline{\text{Mn-O}}}$ where $\overline{\text{ A-O}}$ and $%
\overline{\text{Mn-O}}$ are equilibrium metal-oxygen bond lengths for,
respectively, twelve-fold and sixfold oxygen coordination and its physical
meaning is quite clear. The more colinear the three atoms are, the larger
the transfer integral for charge carriers between them. This is a static,
average property of the system. Temperature-$tf$ phase diagrams as well as
bandwidth effects have been discussed with detail in the bibliography \cite
{hwang2,good,yoshi}, although Hwang {\it et al}. seem to have used incorrect
(ninefold) coordination numbers. Band filling is determined by the doping
level in the system ($x$), also quite relevant. Because of the Coulomb
repulsion in the paramagnetic state, there is a strong tendency of the
system towards charge order as $x$ increases. There is always in the system
an underlying competition between charge order and metallicity, and band
filling is what inclines the scale towards one or the other. Finally, local
disorder effects produced mainly by difference in atomic radii between A$%
^{3+}$ and B$^{2+}$ ions but also by cationic vacancies and oxygen defects
play a very important role. These random defects introduce elastic stress in
the lattice that interferes with the relaxation of the JT effect described
in Fig. 3 affecting the lattice dynamics as well as the hopping process at
high temperatures. The disorder is quantified by means of the variance of
the A-cation radius distribution ($\sigma ^2$) defined as $\sum
y_ir_i^2-\langle r_A\rangle ^2$, where $y_i$ are the fractional occupancies
of the species\cite{rodriguez,sunda}. Rodriguez-Martinez and Attfield find
that the temperature at which the resistivity peaks ($T_p$) and $T_C$ are
monotonically decreasing functions of $\sigma ^2$. At constant $\langle
r_A\rangle $ ( corresponding to the maximum $T_C$ in Fig. 5) they report $%
d(Tp)/d(\sigma ^2)=20600$ K/\AA $^2$ for a series of six samples.

The physical properties of interest for this review are $T_C$, $T_p$, and
lattice transition temperature $T_{latt}$, all determined by the bandwidth,
band filling and local disorder but with different intensity. As a
consequence, while the physical properties are coupled to each other, in
general $T_C\neq Tp\neq T_{latt}$ \cite{ghrao}. It has been reported, on the
other hand, that A-Mn transference is also important, but the experimental
situation is still far from clear\cite{garcia}. Regarding the extrinsic
transport properties of CMR manganites, they are most likely dominated by
grainsize effects, grain boundary scattering and/or spin-flip, irreversible
disorder produced by partial annealing of polycrystalline and film samples,
mechanical strain induced by the substrates\cite{eckstein}, and very
importantly deviations from nominal stoichiometry. Extrinsic effects are
quite important, they are evidently responsible for the largest
magnetoresistance values reported in the literature\cite{chahara} and are
critical for the many prospective technological applications of the
materials. They have been identified as the cause of large low-temperature
magnetoresistance by spin polarized tunneling through intergrain barriers,
as well as anisotropic magnetic effects.

\vspace{1cm}

\begin{center}
HIGH TEMPERATURE TRANSPORT
\end{center}

\vspace{0.5cm}

The transport properties have been, for more that forty years, the easier
and more straightforward characterization and study method for CMR
manganites.\cite{volger} However, not until very recently have the clues in
favor of lattice involvement in the electronic properties been discussed\cite
{tanaka,mj1,mj2}. Fig. 6{\it a} and 6{\it b} display typical results
obtained in polycrystalline samples of composition La$_{2/3}$ Ca$_{1/3}$MnO$%
_3$ prepared by standard solid state reaction techniques. The resistance in
zero field peaks at $T_p\approx 267$ K, somewhat above $T_C$ (H 
\mbox{$<$}%
1Oe) $\approx $ 261 K. The magnetoresistance peaks at $T_p$, but does not
vanish at low temperatures as a consequence of the granular nature of the
specimen. The granular behavior is not relevant in the high temperature
region because the mean free path for charge carriers much smaller that the
grain size; an external magnetic field then has little effect on the
transport properties. A large (intra-grain) mean free path in the metallic
phase below $T_C$, on the other hand, makes the transport extremely
sensitive to intergrain barriers caused by magnetic misalignment between
weakly coupled grains. An external magnetic field reduces those barriers by
aligning neighboring grains. At high temperatures, the thermoelectric power
or Seebeck effect $S(T)$ is also sensitive to the metal-insulator phase
transition. A sharp change from semiconductor-like absolute values $%
|S|\approx 10$ $\mu $V/K toward metallic values $|S|\approx 1$ $\mu $V/K is
found coincident with $T_C$, in Fig 6b). That the Seebeck coefficient
approaches a negative value at high temperatures has been attributed in part
to the reduction in spin entropy produced when a hole converts a Mn$^{3+}$
ion to Mn$^{4+}$ and is not in disagreement with hole doping in the system.
The Seebeck effect, a zero current experiment, shows no indication of grain
boundary effects in the proximity or above $T_C$. These effects are present
at much lower temperatures, where the phonon mean free path approaches the
grain size, as a large spike centered at 30-50 K\cite{mj1,mahendiran}.

The thermopower of semiconductors differ from that of metals also in its
temperature dependence, since is governed by thermal activation of carriers
thus increasing with decreasing temperature:

\begin{equation}
S=\frac{k_B}e\left( \frac{E_S}{k_BT}+B\right)  \label{eqn1}
\end{equation}

In the case of band semiconductors $E_S=E_\sigma $ is the semiconductor gap
defined from the temperature dependence of the conductivity by $\sigma
(T)\propto exp(-E_\sigma /k_BT)$. This is clearly not the case in
manganites, where it has been found that $E_\sigma $ (\symbol{126}100 meV) $%
>>E_S$ (\symbol{126}4 meV). The relatively large activation energy in the
electrical conductivity has to be interpreted in a different way. $E_\sigma $
is, then, not just the semiconducting gap but the gap added to the ``hopping
energy'' $W_H$, a consequence of a thermally activated mobility of localized
carriers jumping between neighboring sites.

The formation and transport properties of small lattice polarons in strong
electron-phonon coupled systems, in which charge carriers are susceptible to
self-localization in energetically favorable lattice distortions, were first
discussed in disordered materials \cite{holstein} and later extended to
crystals \cite{mott}. Emin \cite{emin1} considered the nature of lattice
polarons in magnetic semiconductors, where magnetic polarons are carriers
self-localized by lattice distortions but also dressed with a magnetic
cloud. A transition from large to small polaron occurs as the ferromagnet
disorders, successfully explaining the metal-insulator transition observed
experimentally in EuO. If the carrier together with its associated
crystalline distortion is comparable in size to the cell parameter, the
object is called a small, or Holstein, polaron (HP). Because a number of
sites in the crystal lattice can be energetically equivalent, a band of
localized states can form. These energy bands are extremely narrow, and the
carrier mobility associated with them is predominant only at very low
temperatures. It is important to note that these are not extended states
even at the highest temperatures, where the dominant mechanism is thermally
activated hopping, with an activated mobility $\mu
_p=[x(x-1)ea^2/h](T_0/T)^sexp[-(W_H-J^{3-2s})/k_BT]$ where $a$ is the
hopping distance, $J$ the transfer integral, $x$ the polaron concentration,
and $W_H$ one-half of the polaron formation energy. In the non-adiabatic
limit, we have $s=3/2$ and $k_BT_0=(pJ^4/4W_H)^{1/3}$ and, in the adiabatic
limit, $s=1$ and $k_BT_0=\hbar \omega _0$, where $\omega _0$ is the optical
phonon frequency. The criterion for non-adiabatic behavior is that the
experimental $k_BT_0<<\hbar \omega _0$. Using experimental values for $%
\sigma $, $E_\sigma $, $S$ and cell parameter we find that $k_BT_0/\hbar
\approx 10^{14}s^{-1}$, comparable to optical phonon frequencies, although
it could be considered a marginal case. We will assume the adiabatic limit
to hold, in which case the electrical conductivity, $\sigma =eN\mu _p$ ,
where $N$ is the equilibrium polaron number at a given temperature, can be
expressed as

\begin{equation}
\sigma =\frac{x(1-x)e^2T_0}{\hbar aT}\exp \left( -\frac{\epsilon _0+W_H-J} {%
k_BT}\right)  \label{eqn2}
\end{equation}

Figure 7 shows the same resistance data of figure 6{\it a} displayed in two
different charge localization scenarios, {\it i.e}. adiabatic small
polaron-like: $\log (R/T)$ vs $T^{-1}$ and Variable Range Hopping-like: $%
\log (R)$ vs $T^{-1/4}$. While the data mimics VRH behavior at temperatures
close enough to $T_C$, the adiabatic small polaron (Eq. 2) describes the
system well in the entire temperature range. $R(T)$ data obtained up to $%
T\approx 5T_C$ show excellent agreement with this model\cite
{worledge,worledge2}, and we can assume that the localization of holes
persists up to the material's melting point.

The field dependence of the activation energies $E_S$ and $E_\sigma $ has
been discussed for film samples of similar nominal composition \cite{mj2}.
Within experimental resolution, changes in activation energies are different
but of the same order of magnitude. An estimation of average experimental
values is $\Delta W_H/\Delta H=-2.9\times 10^{-5}$ m{\it e} V/Oe or $%
2.8\times 10^{-5}$ $\%$/Oe and $\Delta E_S/\Delta H=\Delta \varepsilon
_0/\Delta H=-1.4\times 10^{-5}$ m{\it e}V/Oe or $-1.4\times 10^{-4}$ $\%$%
/Oe. While $\varepsilon _0$ reflects changes in the Fermi energy that can be
related to the reported magnetostriction of CMR materials \cite{ibarra},
changes in $W_H$ imply an increase of the radius of the small polaron with
field and consequently some magnetic character of the quasiparticle. Because
of this double character of the localized holes, elastic as well as magnetic
at temperatures up to 2$T_C$, they are named {\em magnetoelastic polarons}
in an attempt to emphasize differences with purely magnetic polarons\cite
{kasuya,mott} and HP.

In Eq. 1, the temperature independent term $B<0$ is given by two
contributions, namely $-(k_B/e)\ln \{[2x(3/2)+1]/[2x^2+1]\}=-(k_B/e)\ln
(4/5)=-19$ $\mu $V/K associated with the spin entropy appropriate for a
spin-1/2 hole moving in a spin-2 background\cite{rafaelle}; and a mixing
entropy term that counts in how many different ways $x$ holes can be
distributed between $n$ Mn sites. In the case of correlated hopping with
weak near-neighbor repulsion\cite{chaikin} this term is $\ln
[x(1-x)/(1-2x)^2]$ and at the nominal doping level $x=1/3$ contributes -60 $%
\mu $V/K; without the repulsive interaction, the mixing term contributes +60
mV/K at the same hole concentration. In either case the prediction is unable
to reproduce the experimental value $S_\infty $ $\approx $ -25 $\mu $V/K,
see Fig. 6{\it b}. Attempts to understand the high temperature limit of the
thermopower $B=S_\infty $ following its changes with hole concentration via
Ca concentration changes have been frustrating so far. An alternative way to
modify the doping level is via control of the concentration of oxygen
vacancies, which can be accomplished with thermogravimetric methods. As part
of a cooperative program with the group at the Centro At\'{o}mico Bariloche,
Argentina\cite{prado} polycrystalline samples were placed in an oven
equipped with atmosphere control capabilities where the isotherm displayed
in Fig. 8 was obtained\cite{caneiro}. The maximum concentration of vacancies
observed without mass instability effects characteristic of phase
segregation was $d=0.051$, enough to depress $T_C$ from 265 K to 221 K.
Figure 9{\it a} shows $S(T)$ vs $1000/T$ for three polycrystalline samples
of composition La$_{0.67}$Ca$_{0.33}$MnO$_3$, La$_{0.67}$Ca$_{0.33}$MnO$%
_{2.49}$, and La$_{0.75}$Ca$_{0.25}$MnO$_3$. Indeed, the variation in $%
S_\infty $ confirms changes in the doping level. The overall changes
however, are not as large as they would be expected. Fig. 9{\it b} show data
by Mahendiran{\it \ et al}.\cite{mahendiran} as well as our samples and
different model predictions. Besides the Chaikin-Beni (Ch-B) model discussed
above, alternative models considered in the bibliography are the correlated
and uncorrelated limits by Heikes\cite{heikes}:

\begin{equation}
S_{ME}^{H_{corr}}=\ln \left( \frac{1+x}{1-x}\right) \text{; }
S_{ME}^{H_{uncorr}}=\ln \left( \frac x{1-x}\right)  \label{eqn3}
\end{equation}

\noindent and the D-dimensional extension of Ch-B formula\cite{rojo}:

\begin{equation}
S_{ME}^D=\ln \left\{ \frac{x(1-Dx)^D}{2[1-x(D+1)]^{D+1}}\right\}
\label{eqn4}
\end{equation}

These predictions are in general unable to reproduce results, except perhaps
in the case of Heikes uncorrelated limit. There are three possible
explanations for this behavior. A disproportionation theory, where two Mn$%
^{3+}$ atoms generate Mn$^{2+}$ and Mn$^{4+}$ sites with the transference of
one electron. The disproportionation density is related to oxygen
non-stoichiometry\cite{hundley}. In an alternative explanation by Emin\cite
{mj3}, the small polarons are proposed to be correlated with divalent atoms
in real space due to the elastic stress introduced in the lattice by atomic
size mismatch. In this kind of ``impurity'' conduction, the number of
available sites for localized states increases with doping and as a
consequence the mixing entropy remains unchanged. Finally, the Heikes
uncorrelated limit suggest that multiple occupancy or collective behavior
could be possible for small polarons. This possibility, unlikely in
principle, finds some support in recent neutron diffraction data where spin
clusters (charge droplets) of a few holes localized in regions of \symbol{%
126 }20\AA\ were identified\cite{neutron}.

Perhaps the most distinctive property of steady-state small-polaronic
transport is its Hall mobility $\mu _{H}$. The activation energy of the Hall
mobility is calculated to be always less than that for drift mobility $E_{d}$
. The simplest model predicts $\approx E_{d}/3$, and this has been observed 
\cite{nagels} before in, for example, oxygen-deficient LiNbO$_{3}$. The sign
of the Hall effect for small polaron hopping can be ``{\it anomalous}.'' A
small polaron based on an electron can be deflected in a magnetic field as
if it were positively charged and, conversely, a hole-based polaron can be
deflected in the sense of a free electron. As first pointed out by Friedman
and Holstein, the Hall effect in hopping conduction arises from interference
effects of nearest neighbor hops along paths that define an Aharonov-Bohm
loop. Sign anomalies arise when the loops involve an odd number of sites.
The first successful measurement of the high-temperature Hall coefficient in
manganite samples was reported by Jaime{\it \ et al}.\cite{mj3}, finding
that it exhibits Arrhenius behavior and a sign anomaly relative to both the
nominal doping and the thermoelectric power. The results are discussed in
terms of an extension of the Emin-Holstein (EH) theory of the Hall mobility
in the adiabatic limit.

The authors exploit the sensitivity of CMR materials to rare-earth
substitutions to lower the transition temperature in (La$_{1-y}$R$_{y}$) $%
_{0.67}$Ca$_{0.33}$MnO$_{3}$ from \symbol{126}260 K at $y=0$ to \symbol{126}
130 K, thereby extending the accessible temperature range to $\approx 4T_{C}$
. The samples used in this study were laser ablated from ceramic targets and
deposited on LaAlO$_{3}$ substrates, as described previously\cite{mj1},
showing a temperature dependence in resistivity and thermopower data
qualitatively similar to Fig. 6.

Sections of these specimens were patterned by conventional lithographic
methods into a five-terminal Hall geometry. Hall experiments were carried
out in a high-temperature insert constructed for use at the 20 Tesla
superconducting magnet at the National High Magnetic Field Laboratory (Los
Alamos, NM). The transverse voltage data taken while sweeping the field from
-16 Tesla to +16 Tesla, and that taken while sweeping back to -16 Tesla,
were each fit to a second-order polynomial with the term linear in field
attributed to the Hall effect. We verified in each case that the
longitudinal magnetoresistance is completely symmetric in field. Figure 10
shows the Hall coefficient derived from the linear term. Several points for
the $y=0$ film are included. Due to the much higher $T_C$ of that sample,
extraction of the Hall contribution leads to greater uncertainty. The data
are, however, consistent with the Gd-substituted film. The line through the
data points is an Arrhenius fit, giving the expression $R_H=-(3.8\times
10^{-11}$ m$^3$/C) $\exp (91$ m$e$V/$k_BT$). Note that the sign is negative,
even though divalent dopants should introduce holes. This is shown more
clearly in an Arrhenius-like plot (inset {\it b}). Inset {\it a}, displays $%
1/R_H$ vs $T$. If the observed linear term in the Hall data is due to the
well known {\it skew} scattering, then $R_H$ is expected to be proportional
to the magnetization and in consequence the data should extrapolate to $T_C$%
. Our data extrapolates to 245 K, more than a hundred degrees above $T_C$. 
{\it Skew} scattering is then unlikely to explain the negative sign of the
Hall data in these samples.

Detailed expressions for the Hall effect in the adiabatic limit have been
calculated by EH\cite{emin3} for the hopping of electrons with positive
transfer integral $J_H$ on a triangular lattice, and results in a normal
(electron-like) Hall coefficient. However, the sign of both the carrier and
the transfer integral changes for hole conduction, leaving the sign of the
Hall coefficient electron-like, and therefore anomalous. However, no anomaly
would arise if the hopping involves 4-sided loops with vertices on
nearest-neighbor Mn atoms. A sign anomaly, then, implies that hopping
involves odd-membered Aharonov-Bohm loops. Such processes arise when
next-nearest neighbor ({\em nnn}) transfer processes across cell face
diagonals are permitted. If the Mn-O-Mn bonds were strictly colinear, the
former processes would be disallowed by symmetry. However, the bond angles
are substantially less than 180$^{\circ }$, implying the presence of $p$
-bond admixture, and opening a channel for diagonal hops. The
triangular-lattice calculation of EH is extended to the situation in which a
hole on a Mn ion can hop to any of its four nearest neighbors in the plane
normal to the applied field with transfer matrix element $J<0$ and to its
four {\em nnn} with a reduced transfer energy $\gamma J$. The effect of
these diagonal hops (plus those in the plane containing both electric and
magnetic fields) has also to be considered on the conductivity prefactor.
The Hall coefficient can be written as $R_H=R_H^0(T)\exp (2E_\sigma /3k_BT)$%
, with

\begin{equation}
R_H^0=-\frac{g_H}{g_d}\frac{F(|J_H|/k_BT)}{ne}\exp \left\{ -[\varepsilon
_0+(4|JH|-E_S)/3]/k_BT\right\}  \label{eqn5}
\end{equation}

EH found that the factor $g_{H}=1/2$ for three-site hopping on a triangular
lattice. In Eq. 7 the carrier-density is included as $n\exp (-E_{S}/k_{B}T)$
, where $E_{S}$ is estimated to be 8 m$e$V from the thermopower data. The
quantity $\varepsilon _{0}$ is the $J_{H}$-dependent portion of a carrier's
energy achieved when the local electronic energies of the three sites
involved in an Aharonov-Bohm loop are equal. For the problem considered by
EH, an electron hopping within a lattice composed of equilateral triangles, $%
\varepsilon _{0}=-2|J_{H}|$, and $g_{H}/g_{d}=1/3$. Within the domain of
validity of EH, the temperature dependence of $R_{H}$ arises primarily from
the factor $\exp (2E_{\sigma }/3k_{B}T)$ when $E_{\sigma }>>E_{S}$. For
holes hopping within a cubic lattice in which three-legged Aharonov-Bohm
loops include $\varepsilon _{0}$ varies from $-\sqrt{2}|J_{H}|$ to $-|J_{H}|$
as $\gamma $ increases from zero to unity, and the temperature dependence of 
$R_{H}$ remains dominated by the factor $\exp (2E_{\sigma }/3k_{B}T)$.
Indeed, the energy characterizing the exponential rise of the Hall
coefficient that we observe, $E_{H}=91\pm 5$ m$e$V, is about 2/3 the
measured conductivity activation energy, $E_{H}/E_{\sigma }=0.64\pm 0.03$,
in excellent agreement with theory.

The geometrical factor $g_d$ depends on the ratio of the probability P{\em %
nnn} of {\em nnn} hops to P{\em nn}, that of {\it nn} hops, through $%
g_d=(1+4Pnnn/Pnn)$. If these probabilities are comparable $(\gamma \approx
1) $ $g_d=5$, $g_H=2/5$ and the exponential factor in Eq. 7 becomes $\exp
[(E_S-|J_H|)/3k_BT]\approx 1$. In the regime $|J_H|\geq k_BT$, the function $%
F(|J_H|/k_BT)$ is relatively constant with a value $\approx 0.2$, and we
find $R_H^0\approx -0.02/ne=-3.8x10^{-11}$ m$^3$/C. This yields an estimated
carrier density $n=3.3\times 10^{27}$ m$^{-3}$, quite close to the nominal
level of $5.6\times 10^{27}$ m$^{-3}$.

Before moving over to the low temperature transport properties the following
conclusions can be reached. The high-temperature Hall coefficient in
manganite films is consistent with small-polaron charge carriers that move
by hopping. The magnitude of the conductivity prefactor indicates that the
carrier motion is adiabatic. The sign anomaly in the Hall effect implies
that small polarons hop not only among near-neighbor sites (making
Aharonov-Bohm loops with an even number of legs) but must have a significant
probability of traversing Hall-effect loops with odd numbers of legs. As
such, the results indicate the occurrence of significant {\em nnn} transfer
across face diagonals, and therefore a crucial role for deviations of the
Mn-O-Mn bond angle from $180^{\circ }$. In other words, the sign anomaly its
a simple consequence of the geometry of the sublattice where the small
polarons move and the fact that it is triangular and not square indicate an
interesting possibility, that may also relate to unusual high-temperature
values observed for the Seebeck coefficient \cite{mj1,mj2}. That is that
transport is a type of impurity conduction in which carriers remain adjacent
to divalent cation dopants ({\it i.e}. Ca ions). The local distortions
associated with the presence of the impurity may also increase the admixture
of $\pi $-bonds, and enhance diagonal hopping.

In a recent paper, Worledge {\it et al.}\cite{worledge2} discuss the
temperature and doping dependence of the resistivity in La$_{1-x}$Ca$_x$MnO$%
_3$ laser ablated films measured up to $T=1200$ K for $0>x>1$. They conclude
that the results can be unambiguously explained by adiabatic small polaron
hopping, which is limited by on-site Coulomb repulsion. The magnitude of the
conductivity prefactor, however, is too large to be accounted by the
classical theory by Emin and Holstein\cite{emin3} and the authors claim that
a proper description should consider hopping beyond nearest neighbors, in
good agreement with high temperature Hall effect results. A few other
reports on the Hall effect of manganites are now available\cite
{earlyhall,hall}, they are restricted however to the relatively low
temperature side of the diagram in Fig.4,{\it \ i.e}. regions I and III. The
low temperature Hall effect is not less intriguing than the high temperature
counterpart, and is not yet understood. The Hall resistance in the metallic
regime imply carrier concentrations up to 3$\times $ the nominal values
suggesting some compensation effects and/or two-band conduction, the
spontaneous Hall contribution is opposite in sign from the normal Hall
effect with the overall effect exhibiting a sign change around $T_C$ from
hole-like in the ferromagnetic phase to electron-like in the paramagnetic
phase. In order to clarify the subject, more experiments in the very high
temperature limit ($T\geq 2T_C$) are desirable..

\vspace{1cm}

\begin{center}
LOW TEMPERATURE TRANSPORT
\end{center}

\vspace{0.5cm}

The low temperature region I, in Fig. 4 in optimally doped cubic manganites
is perhaps the most interesting one since it corresponds to a ground state
that is the closest to half metallic systems ever synthesized.
Unfortunately, the transport properties of polycrystalline samples in this
regime are dominated, or at best highly influenced, by grain boundary
scattering, and close attention has to be paid to sample quality issues.
These problems have led in the past to misunderstanding about the most basic
transport properties, like resistivity and Seebeck effect for example. One
common problem in the resistivity of polycrystalline specimens is the
presence of a minimum at $T\approx $10-20 K that has been attributed to
localization effects. In the same temperature range, the Seebeck effect
shows large anomalies (as big as -40 $\mu $V/K) by no means compatible with
a metallic state. None of these features have been reproduced in carefully
prepared ($x\sim $1/3) single crystals and are thus considered
non-intrinsic. One of the simplest methods of characterization seems to be
the magnetoresistance (MR), since granular samples show large ratios in
small applied magnetic fields down to the lowest temperatures. On the other
hand, MR rapidly vanishes below $T_C$ in long time annealed films\cite{mj1}
and single crystals.

The double exchange mechanism is generally agreed to provide a good
description of the ferromagnetic ground state. In that model, strong Hund's
Rule coupling enhances the hopping of $e_g$ electrons between neighboring Mn$%
^{3+}$ and Mn$^{4+}$ ions by a factor $\cos (\theta /2)$, where $\theta $ is
the angle between the spin of their respective $t_{2g}$ cores, thereby
producing a ferromagnetic interaction. In KO's treatment of the problem\cite
{kubo}, occupied sites are assumed to have total spin $S=2$, the combination
of the spin-3/2 $t_{2g}$ core and spin-1/2 $e_g$ electron demanded by strong
Hund's rule exchange. Holes are then assumed to couple antiparallel to each
localized spin. In the ferromagnetic ground state, with all local spins
aligned, only spin-down holes can move to form a band. However, once the
system begins to disorder, a locally down-spin hole is the appropriate
combination of majority and minority carriers, referred to the global
magnetization axis. Therefore, the minority-spin hole band reappears as the
system disorders, even though the local moments retain their Hund's rule
value. Furukawa has treated this explicitly in a many-body context,
demonstrating that both minority and majority bands are split by Hund's rule
exchange in the paramagnetic state. As the system magnetizes, the lower
majority-spin band gradually gains spectral weight at the expense of the
lower minority-spin band . Both treatments predict that the ground state is
half-metallic; that is, that the carriers are fully spin-polarized. We can
leave aside here electron-phonon coupling which dominates near and above $%
T_C $.

At low temperature, there are no propagating minority-spin hole states in
the $S=2$ manifold; they only exist on sites at which the $t_{2g}$ core is
not ferromagnetically aligned. As a consequence, single-magnon scattering
processes, which cause the resistivity of conventional ferromagnets to vary
as $T^2$, are suppressed. KO extended the standard perturbation calculation
of Mannari\cite{mannari} to consider two-magnon processes, predicting a
leading $T^{9/2}$ temperature dependence of the resistivity. However, a
dominant $T^2$ contribution is universally observed in the manganites, and
has usually been ascribed to electron-electron scattering.\cite
{urushi,schiffer,mj1} New resistivity data on single crystals is discussed
below, demonstrating that the quadratic temperature dependence is strongly
suppressed as the temperature is reduced. The constancy of the
low-temperature resistivity has been noted elsewhere, but not explained.\cite
{snyder} It is argued that the observed $T^2$ contribution reflects the
reappearance of minority spin states that are accessible to thermally
excited magnons. Quite recently, spin-polarized photoemission data, taken on
films exhibiting square hysteresis loops, indicate 100\% spin polarization
at low temperatures, decreasing as the temperature is increased\cite{park}.
Single crystals, which have essentially no hysteresis, would be expected to
depolarize more rapidly. To explore the consequences, Mannari's calculation
is extended to the situation in which a minimum magnon energy is required to
induce spin-flip transitions. At temperatures well below that energy, single
magnon scattering is suppressed exponentially as predicted by KO. The
treatment discussed by Jaime{\it \ et al}.\cite{mj4} is in the context of
the relaxation time approximation while a proper theory would consider
lifetime effects from magnon scattering using Furukawa's many-body approach.
Nonetheless, the results are in qualitative agreement with the data. Band
structure calculations also indicate that minority spin states persist at $%
E_F$, even at $T=0$ K.\cite{pickett}

High quality single crystals of nominal composition La$_{0.66}$(Pb$_{0.67}$Ca%
$_{0.33}$)$_{0.34}$MnO$_3$, determined by inductively coupled plasma
spectroscopy on samples from the same batch, were grown from a 50/50 PbF$_2$%
/PbO flux and used to study the low temperature properties. X-ray
diffractometry shows a single pseudo-orthorhombic structure with lattice
parameters $a$ = 5.472(4) \AA , $b$ = 5.526(6) \AA , and $c$ = 7.794(8) \AA
. Gold pads were evaporated onto both oriented and unoriented crystals using
both standard four-terminal and Montgomery eight-corner contact arrangements
as described elsewhere\cite{mj4}. Fig. 11 shows the resistivity of sample
sc3, a single crystal of dimensions $1.04\times 1.24\times 0.3$ mm$^3$ with $%
T_C$ = 300 K, vs the square of the temperature in fields up to 70 kOe. The
data show a dominant $T^2$ temperature dependence with evidence of a small $%
T^5$ contribution (10 $\mu \Omega $cm at 100 K). A calculation of the $%
T^{9/2}$ contribution predicted by KO for two-magnon processes predicts only
0.5 $\mu \Omega $cm at 100 K with appropriate parameters. It is likely,
then, that this is the usual $T^5$ contribution from electron-phonon
processes. Within the spin-wave approximation, the low temperature
magnetization is given by $M(T)=M(0)-BT^{3/2}-...,$ where $B=0.0587g\mu
_B(k_B/D)^{3/2}.$ The stiffness constant $D$ has been determined by neutron
scattering \cite{lynn,baca} and muon spin resonance \cite{heffner} to be $%
D\approx 135-170$\ meV \AA $^2.$ The right side of Fig. 11 shows the
magnetization for this sample, from which we extract $B(10\;$kOe$)$ and the
value $D=165$ meV \AA $^2,$ in good agreement with other results.

The plot in the right on Fig. 12 shows that the data do not follow a $T^2$
dependence to the lowest temperatures. Rather, they deviate gradually from
the curve $\rho _0$ + $\alpha (H)T^2$, fit over the range 60 $\leq T\leq $%
160 K, saturating at an experimental residual resistivity $\rho _0^{exp}$ =
91.4 $\mu \Omega $cm, comparable to values observed by Urushibara et al.\cite
{urushi}, but \symbol{126}7\% larger than $\rho _0$. This conclusion is not
changed by including the $T^5$ contribution. Fits to data taken in various
fields show that $\alpha (H)$ decreases with increasing field and is the
source of the small negative magnetoresistance at low temperatures. To
quantify the disappearance of the $T^2$ contribution, the authors calculate $%
(\rho -\rho _0^{exp})/\alpha (H)T^2$ and display it in Fig. 13{\it a}. The $%
T^5$ contribution which gives a slight upward curvature to the data at
higher temperatures has not been substracted. Should the $T^2$ description
be valid in the low temperature range, we should have $(\rho -\rho
_0^{exp})/\alpha (H)T^2$ $\equiv 1$. Note that this description of the data
is extremely sensitive to the value of $\rho _0^{exp}$, and it must be
determined very carefully. An alternative description is possible by means
of a numerical derivative, as discussed elsewhere\cite{mj4}, similar
conclusions are arrived at.

Previous investigators have attributed the $T^2$ term in the resistivity to
electron-electron scattering. An empirical relationship has been found
between the coefficient $\alpha $ and the coefficient $\gamma _{hc}$ of the
electronic specific heat by Kadowaki and Woods\cite{kadowaki}: $\alpha
/\gamma ^2=1\times 10^{-5}$ $\mu \Omega $cm(mole K/mJ)$^2$. Using our
experimental value and $\gamma $ \symbol{126}4 mJ/mol K$^2$ from ref. \cite
{hamilton} we find a value \symbol{126}60$\times $ the Kadowaki-Woods
parameter which argues against {\it e}-{\it e} scattering. With an electron
density at the nominal doping level $n=5.7\times 10^{27}$ m$^{-3}$, the
effective mass that follows from $\gamma _{hc}$ is $m^{*}/m$ = 2.5 and the
Fermi energy is $E_F=0.5$ $e$V and the $e-e$ relaxation rate of the order of 
$2\times 10^{11}$ $s^{-1}$ at 100 K. The experimentally observed $T^2$
contribution at that temperature is 100 $\mu \Omega $ which with the same
parameters correspond to a relaxation rate of $6\times 10^{13}$ $s^{-1}$
more that two orders of magnitude larger. This disagreement is not fixed by
using low temperature Hall-deduced effective concentration of carriers\cite
{hall}. Rather than vanishing, what is more, $e-e$ scattering should become
more apparent as the temperature is reduced. We conclude that $e-e$
scattering is an unlikely explanation for the observed quadratic dependence
on temperature.

When the usual calculation of the electron-magnon resistivity\cite{mannari}
is extended to allow the minority-spin sub-band to be shifted upward in
energy such that its Fermi momentum differs by an amount $q_{min}$ from that
of the majority sub-band, the one-magnon contribution can be written as $%
\rho _\epsilon =\alpha _\epsilon T^2$, where $\alpha _\epsilon =(9\pi
^3N^2J^2\hbar ^5/8e^2E_F^4k_F)(k_B/m^{*}D)^2I(\epsilon )$. Here $NJ$ is the
electron-magnon coupling energy which is large and equal to $\mu =W-E_F$ in
the DE Hamiltonian of KO; $2W$ is the bandwidth.. The magnon energy is given
by $Dq^2$, and

\begin{equation}
I(\epsilon )=\int\limits_{{\epsilon }}^{{\infty }}\frac{x^2}{sinh^2x}dx
\label{eqn6}
\end{equation}

\noindent The lower limit is $\epsilon =Dq_{min}^2/2k_BT$, where $Dq_{min}^2$
is the minimum magnon energy that connects up- and down-spin bands; result
that reproduces Mannari's calculation in the limit $\epsilon \rightarrow 0$,
and KO's exponential cut-off for large $\epsilon $. At high temperatures,
the lower limit of the integral in Eq. 6 can be set equal to zero, leaving
only the coupling energy $NJ=W-E_F$ as a parameter. Equating the calculated
value to the experimental $\alpha $ fixes the coupling to be $W-E_F$ $%
\approx 1.0$ eV or $W\simeq 1.5$ eV, in good agreement with a virtual
crystal estimate of the band width. \cite{pickett} In Fig. 12 we have
plotted $I(\epsilon ,T)$ assuming $D(0)q_{min}^2=4$ meV and including the
temperature dependence observed experimentally, $D(T)/D(0)=(1-T/T_C)^{0.38}$ 
\cite{baca} which is important only at higher temperatures. While the curve
follows the data qualitatively, it is clear that the minimum magnon energy
is substantially larger than 4 meV at low temperatures, and decreases
rapidly with increasing temperature.

Fig. 13 shows the Seebeck coefficient $S(T)$, measured on the same
unoriented sample. Below $20$ K, $S(T)$ is positive as expected for hole
conduction, linear in temperature, and extrapolates to zero as $T$ $%
\rightarrow 0$. If we take the scattering to be independent of energy, which
is the case below $20$ K, Seebeck coefficient can be expressed as $S(T)=(\pi
^2/2e)(k_B^2T/E_F).$\cite{am} Using the simplistic approximation of
parabolic band $E_F=\hbar ^2k_F^2/2m^{*}$, and spherical Fermi surface $%
k_F^3=3n\pi ^2$, the effective mass results to be $m^{*}/m\thickapprox 3.7$,
comparable to the value obtained from specific heat measurements. The sharp
deviation from linear behavior in the temperature range $20-40$ K correlates
with the onset of electron-magnon scattering which, being a spin flip
process, must involve the minority spin band, and which therefore has a
different dependence on energy near $E_F$.

In conclusion, the low temperature transport data cannot be explained by
electron-electron scattering as proposed before\cite
{mj1,urushi,schiffer,snyder} and, while oversimplified, the extension of the
standard calculation of one-magnon resistivity to describe spin-split bands
gives a qualitative account of the half metallic suppression of the
spin-wave scattering at very low temperatures.

\vspace{1cm}

\begin{center}
INTERMEDIATE\ TEMPERATURES, ${\bf T\simeq T}_C$
\end{center}

\vspace{0.5cm}

As Millis and coworkers \cite{millis1,millis2,millis3,millis4} have
emphasized, the Jahn-Teller effect in Mn$^{3+},$ if strong enough, can lead
to polaron formation and the possibility of self-trapping. The effective JT
coupling constant $\lambda _{eff}$, in this picture, must be determined
self-consistently, both because it depends inversely on the bandwidth and
because the effective transition temperature increases with decreasing $%
\lambda _{eff}.$ If $\lambda _{eff}$ is larger than a critical value $%
\lambda _c$, the system consists of polarons in the paramagnetic phase. As
the temperature is lowered to the Curie temperature T$_C,$ the onset of
ferromagnetism increases the effective bandwidth, which reduces $\lambda
_{eff},$ thereby increasing the effective transition temperature. As a
result, the polarons may dissolve into band electrons if $\lambda _{eff}$
drops below $\lambda _c$ and the material reverts to a half-metallic, double
exchange ferromagnet at low temperatures. The tendency toward polaron
formation is monitored by a local {\em displacement} coordinate $r$, which
is zero for $\lambda _{eff}<\lambda _c,$ and grows continuously as $\lambda
_{eff}$ increases beyond that value. However, polarons are typically \cite
{emin3} bimodal--large or small--so that we should consider $r$ to be a
measure of the relative proportion of large polarons (band electrons for
which $r\approx 0$) and small polarons (for which $r$ is an atomic scale
length).

Indeed, there is growing experimental evidence\cite
{louca1,yoon,booth3,billinge} that polaronic distortions, evident in the
paramagnetic state, persist over some temperature range in the ferromagnetic
phase, as displayed in Fig 2 as a coexistence zone. This possibility can be
explored by considering the observed electrical resistivity to arise from
the parallel conduction of a field- and temperature-dependent polaronic
fraction (with activated electrical conductivity) and band-electron fraction
(with metallic conductivity). The validity of this model is tested by
applying it to the thermoelectric coefficient using an extension of the
well-known Nordheim-Gorter rule for parallel conducting channels. The La$%
_{2/3}$Ca$_{1/3}$MnO$_3$ film samples used in this study were prepared by
pulsed laser deposition onto LaAlO$_3$ substrates to a thickness of 0.6 $\mu 
$m$.$ As described previously \cite{mj2}, they were annealed at 1000 ${%
{}^{\circ }}$C for 48 hr. in flowing oxygen. Measurements were carried out
in a 7T Quantum Design Magnetic Property Measurement System with and without
an oven option provided by the manufacturer. A modified sample rod brought
electrical leads and type-E thermocouples to the sample stage. A bifilar
coil of 12 $\mu $m Pt wire was calibrated to serve both as a thermometer and
to provide a small heat input for the thermopower measurements. Measurements
in fields up to 70 kOe could be carried out over the temperature range $4$ K 
$\leq $ $T\leq 500$ K. Following the transport measurements, magnetization
data $M(H,T)$ were acquired up to $380$ K by conventional methods.

Figure 14{\it a} shows the resistivity data in zero field over the full
temperature range. The data below $200$ K exhibit metallic behavior, and are
well fit by a power law, $\rho _{lt}(T)=[0.22+2\times 10^{-5}$ K$%
^{-2}T^2+1.2\times 10^{-12}$ K$^{-5}T^5]$ m$\Omega $ cm. Above 260 K, the
resistivity is exponential, given\cite{mj1} by the form expected for the
adiabatic hopping of small polarons, $\rho _{ht}=(1.4\mu \Omega $ cm K$%
^{-1})T\exp (1276$ K$/T)$. These are shown as broken lines. The assumption
is that these represent the resistivity of band electrons and polarons,
irrespectively, and that the transition region can be represented by a
parallel combination characterized by a mixing factor $c(H,T)$ which is
envisaged to be the fraction of the carriers that are in the metallic state;
that is 
\begin{equation}
\rho (H,T)=\left[ \frac{c(H,T)}{\rho _{lt}(T)}+\frac{1-c(H,T)}{\rho _{ht}(T)}%
\right] ^{-1}  \label{eqn7}
\end{equation}

As a first approximation $c(0,T)=M(0,T)/M_{sat}$ is chosen, using the data
in the inset of Fig. 14{\it a}. The solid curve through the data shows the
result of this process with no further adjustable parameters. As a second
test of this approach, the Seebeck coefficient $S(H,T)$ is considered,
measured over the same temperature range and plotted in Fig. 14{\it b}. We
fit the low temperature thermopower arbitrarily to a power law, $%
S_{lt}(T)=[(0.051$ K$^{-1})T-(1.3\times 10^{-4}$ K$^{-2})T^2-(3.2\times
10^{-7}$ K$^{-3})T^3]$ $\mu $V/K, and the high temperature data \cite{mj2}
to the form expected for small polarons, $S_{ht}(T)=[(9730$ K$)T^{-1}-29]$ $%
\mu $V/K. Broken lines in Fig. 14{\it b} show the extrapolation of these
fits into the transition region. The Nordheim-Gorter rule\cite{bernard} can
now be applied to compute the thermopower for parallel conduction, 
\begin{equation}
S(H,T)=\rho _{exp}(H,T)\left[ \frac{c(H,T)S_{lt}(T)}{\rho _{lt}(T)}+\frac{%
(1-c(H,T))S_{ht}(T)}{\rho _{ht}(T)}\right]  \label{eqn8}
\end{equation}
The result is shown as a solid line in Fig. 14{\it b}, again using the
reduced magnetization as a measure of the relative concentration of band
electrons and polarons.

The association of $c(H,T)$ with $m(H,T)\equiv M(H,T)/M_{sat}$ does not hold
in applied fields. Fig 15{\it a} shows the magnetization in fields up to 50
kOe. In Fig. 15{\it b}, the dashed curve shows the calculated $\rho (10 $
kOe,$T)$ along with the experimental data. Clearly, $m(H,T)$ significantly
overestimates the mixing factor $c(H,T).$ In order to explore this two-fluid
approach further, the mixing coefficient is {\em computed }from the {\em %
field-independent }low and high temperature resistivities, and its validity
tested by calculating from it the field-dependent Seebeck coefficient.
Explicitly, $c(H,T)$ is defined through the expression 
\begin{equation}
c(H,T)=\frac{\rho _{ht}(T)/\rho _{exp}(H,T)-1}{\rho _{ht}(T)/\rho _{lt}(T)-1}
\label{eqn9}
\end{equation}
which clearly approaches zero and unity in the high and low temperature
limits respectively. Fig. 16{\it a} shows the mixing factor at various
applied fields extracted from the data of Fig. 15{\it b}. In Fig. 16{\it b},
these experimental mixing factors are used in Eq. 8 to generate curves for
the field dependent Seebeck coefficient. These give an excellent account of
the data, providing an independent check on the validity of this two-fluid
approach. The main effect of the magnetic field is to shift the onset of the
band-electron phase without broadening the transition. However, as we shall
see, the vanishing of $c(H,T)$ does not represent a shifted critical point
for the material.

The essential feature of the Millis {\it et al.} model is that the effective
Jahn-Teller coupling constant is very near its critical value in the
paramagnetic phase. In this case, coupling to the magnetization via the
associated band-broadening of the double exchange model, reduces $\lambda
_{eff}$ through its critical value $\lambda _c,$ inducing the expansion of
small polarons into band electrons. A simple mean-field model is proposed
here, that reproduces the essential features of the microscopic model and
provides a comparison with experiment. The assumed ferromagnetic free-energy
functional is of conventional form 
\begin{equation}
F_{mag}=\frac 12(T/T_C-1)m^2+\frac 14bm^4-mh  \label{eqn10}
\end{equation}

\noindent where the free energy is written in units of \cite{mattis} $%
3Sk_BT_C/(S+1)=1.94k_BT_C$ for $S=2(1-x)+3x/2=1.83$ and $x=1/3$, and $h=g\mu
_B(S+1)H/3k_BT_C=H/2360$ kOe$.$ The dependence of $\lambda _{eff}$ on the
magnetization can be approximated by writing $\lambda _{eff}-\lambda
_c\varpropto \alpha -\gamma m^2+...,$ where $\alpha $ is small and positive.
The electronic free energy can then be written, in the same dimensionless
units as Eq.10, as 
\begin{equation}
F_{el}=\frac 12(\alpha -\gamma m^2)c^2+\frac 14\beta c^4  \label{eqn11}
\end{equation}

\noindent Here $c(H,T)$ is a nearly-critical secondary order parameter,
driven by the difference $\lambda _{eff}-\lambda _c.$ Minimizing the total
free energy two coupled equations are obtained, $(T/T_C-1-\gamma
c^2)m+bm^3-h=0$ and $(\alpha -\gamma m^2)c+\beta c^3=0$. From the later it
is obvious that the concentration of metallic electrons is zero until the
magnetization reaches the value $m=\sqrt{\alpha /\gamma },$ beyond which
point $c$ increases. In the limit $\alpha \rightarrow 0,$ $c$ is
proportional to $m$, with the result that $b\rightarrow b-\gamma ^2/\beta ,$
signalling a tendency for the system to approach a tricritical point and
first order transitions as the coupling constant is increased. Note that the
existence of a non-zero concentration $\overline{c}$ can be considered to
increase the critical point to $(1+\gamma \overline{c}^2)T_C,$ causing the
magnetization to increase more rapidly than would be the case without
coupling to the metallic electron concentration. Solutions to the coupled
equations are, 
\begin{equation}
m=B_S\left( \frac{3ST_C}{(S+1)T}[(1+\gamma c^2)m+h)\right)  \label{eqn12}
\end{equation}

\noindent and 
\begin{equation}
c=\tanh \left[ (1-\alpha +\gamma m^2)c\right]  \label{eqn13}
\end{equation}

In Fig. 17 the simultaneous solutions of Eqs. 12 \& 13 for $\alpha =0.02$
and $\gamma =0.3$ at $H=0,$ $24$ kOe, and $48$ kOe is found. Application of
the magnetic field increases the temperature at which $c$ becomes non-zero
by 7\% or 20 K, consistent with the experimental data in Fig. 16{\it a}, but
does not produce a high-temperature tail. As no thermal factors are included
in the definition of $c,$ the concentration of free carriers does not
approach unity, and therefore differs slightly from the experimentally
defined $c(H,T)$ in Eq. 9. The abrupt appearance of band electrons in this
model produces a kink in the zero field magnetization curve at the onset
temperature $T_D$, seen as a deviation from the $H=0$, $\gamma =0$ curve.

In non-zero field, the kink persists as seen in Fig. 18{\it a} where we plot 
$\chi ^{-1}\equiv H/m$ at several magnetic fields. These results show
clearly how the delocalization of charge carriers produces a rise in the
effective $T_C$, in good agreement with experimental data\cite{goodwing} for
La$_{0.79}$Ca$_{0.21}$MnO$_3$. The magnitude of the kink present in the
experimental $\chi ^{-1}(T)$ is larger and more evident in samples that show
broader ferromagnetic transitions at constant doping,\cite{jason} and are
consequently considered of ''lower quality''. This deserve a further
analysis. Fig. 18{\it b} shows the resistivity curves determined using the
mean-field $c(H,T)$. Clearly, the model must be extended to include critical
fluctuations and associated rounding.

The proposed model differs from a percolation-like picture in which more or
less static regions of high conductivity are weakly connected by surrounding
insulating material. If that were the case, the standard Nordheim-Gorter
rule for series connection would emphasize the increasing Seebeck
coefficient of the resistive polaronic contribution, rather than the small
thermopower of the more conductive component. There is ample experimental
evidence, from studies of spin waves for example, that the ordered phase
emerges with its full three-dimensional properties ---albeit with strong
evidence of slow, diffusive contributions--- in materials in the composition
regime discussed here. This mean-field model ignores a number of features
that should be included in a complete treatment. In particular a term $m^2c$
is missing, because it leads to a first-order transition for all values of
the parameters; it cannot be ruled out on symmetry grounds. Similarly, there
should be a mixing entropy in the electronic free energy which, at
sufficiently high temperatures, will lead to thermal dissociation of the
polarons. Finally, no gradient terms were included and therefore ignore
inhomogeneous thermal fluctuations that are certain to be significant in a
system such as this where there are competing order parameters. Nonetheless,
this phenomenological approach provides a qualitative understanding of the
field and temperature dependence of the transport properties while correctly
predicting the existence of kinks in the magnetization curves.

In summary, we have discussed the transport properties of optimally doped
manganite materials and showed how they play a key role in the understanding
of their ground state, as well as their different magnetic phases. Transport
properties allow us to distinguish different temperature regimes and also to
identify the relevant physics ruling them. As indicated in Fig. 4, DE
physics dominates the very low temperature region. Additional theoretical
work is needed however to describe the details of the gradual changes in the
band structure with temperature, from 100\% spin-polarized to partial
polarization just bellow $T_C$. Such a model should allow us to properly
calculate the temperature dependence of the electrical resistivity at very
low temperature. JT and localization of charge dominates at high
temperatures, but more experimental work is needed to improve the
understanding of the Seebeck and Hall effects. One of the hardest
experimental problems is related to sample quality issues. As discussed
before, the physical properties of manganites are strongly dependent on
bandwidth, doping, and local defects. Most of the experimental work done
until now has concentrated on samples where these three parameters are
changed simultaneously. For example, studies of $T_C$ vs doping do not
usually take into account the concentration of local defects nor the
tolerance factor. It would be useful for the understanding of the high
temperature Seebeck effect to be able to prepare samples with different
doping levels and different concentration of local defects, keeping the
tolerance factor $tf$ a constant. Samples like these were prepared at
Urbana, to test the impurity conduction model proposed by Emin, however
discrepancy between nominal and measured chemical compositions make the
experimental results hard to analyze\cite{peter}. At intermediate
temperatures, both DE and JT mechanisms are required to understand the
details of the phase transition between a paramagnetic insulator and a
highly polarized ferromagnetic metal. The gradual delocalization of charge
carriers is driven by a temperature dependent effective coupling between
charge and lattice, which at the same time is determined by a DE controlled
bandwidth.. While the coexistence of itinerant and localized charges explain
some experimental properties the situation is still unclear as respects the
origins and mechanisms of very slow spin dynamics and cluster formation just
above $T_C$. More careful measurements in this regime as well as a
theoretical description that includes both double exchange, strong
electron-phonon coupling, and spin fluctuations, should help.

We would like to thank A. Caneiro and F. Prado, Centro At\'{o}mico
Bariloche; M. Rubinstein, U.S. Naval Research Laboratory; D. Emin,
University of New Mexico, Albuquerque, N.M.; P. Han, P. Lin, and S.-H. Chun,
University of Illinois at Urbana. This work was possible thanks to the
support from U.S. Department of Energy at Los Alamos National Laboratory,
NM, USA. MBS acknowledges support by the Department of Energy, Office of
Basic Energy Sciences through Grant No. DEFG0291ER45439 at the University of
Illinois and by National Science Foundation Grant No. DMR-9120000 through
the Science and Technology Center for Superconductivity.

\begin{figure}
\caption{a) Lattice structure of LaMnO$_3$ b) MnO planes showing the characteristic 
periodic distortion c) Mn$^{3+}$ electronic levels $t_{2g}$ and $e_g$ split as a result
 of the Mn-O bond length distortion qualitatively displayed in b)}
\label{ fig1}
\end{figure}%

\begin{figure}
\caption{Phase diagram for A$_x$B$_{1-x}$MnO$_3$ manganites. Modified from 
Schiffer et al. [76], mesh is coexistance region discussed below.}
\label{ fig2}
\end{figure}%

\begin{figure}
\caption{Strain field induced in the structure by a non-JT, s = 3/2 spin, Mn$^{4+}$ 
atom.                                                       }
\label{fig3 }
\end{figure}%

\begin{figure}
\caption{Different relevant temperature ranges for transport properties in CMR 
manganites for doping levels $x\approx $1/3. At low temperatures DE effects are 
dominant, while at high temperature dynamic structural effects control the transport 
properties.}
\label{fig4}
\end{figure}%

\begin{figure}
\caption{Transition temperature vs. tolerance factor for the 
A$_{0.7}$B$_{0.3}$MnO$_3$ family of compounds, modified from ref. 53. PI: 
paramagnetic insulator, PM: paramagnetic metal, CFI: canted ferromagnetic insulator, 
FGC: ferromagnetic glass conductor, FM: ferromagnetic metal, O and O: orthorrombic, 
R: rombohedric phases. The shadowed area indicates coexistence of extended and 
localized electronic states. Some samples discussed bellow were included: 
($\otimes $) (La$_{0.33}$Gd$_{0.33}$)Ca$_{0.33}$MnO$_3$, 
($\square $) (La$_{0.5}$Gd$_{0.17}$)Ca$_{0.33}$MnO$_3$, 
($\bullet $) La$_{0.67}$Ca$_{0.33}$ MnO$_3$, and 
($\blacksquare $) La$_{0.67}$(Ca$_{0.11}$Pb$_{0.22}$)MnO$_3$.}
\label{ fig5}
\end{figure}%

\begin{figure}
\caption{Resistance vs temperature for a polycrystalline sample in zero field 
($\square $) and H=5T ($\bullet $) and exponential fit. The data follows very well 
the model for adiabatic small polarons. b) The thermoelectric power vs temperature 
and fit to a function of the form A/T + B.}
\label{ fig6}
\end{figure}%

\begin{figure}
\caption{The resistance for a polycrystalline sample of composition 
La$_{0.67}$Ca$_{0.33}$ MnO$_3$ ploted in two different scales, the one 
expected for adiabatic small polaron hopping and for Variable Range Hopping.}
\label{fig7 }
\end{figure}%

\begin{figure}
\caption{The 1000 $^{\circ }$C isotherm for La$_{0.67}$Ca$_{0.33}$MnO$_3$.                                                                                           }
\label{fig8 }
\end{figure}%

\begin{figure}
\caption{a) Thermopower vs. 1000/T for polycristalline samples of composition 
La$_{0.67}$ Ca$_{0.33}$MnO$_3$, La$_{0.67}$Ca$_{0.33}$MnO$_{2.49}$ and 
La$_{0.75}$Ca$_{0.25}$MnO$_3$. b) $S_\infty $ vs Mn$^{4+}$ content for our 
samples and theoretical predictions discussed in the text.}
\label{ fig9}
\end{figure}%

\begin{figure}
\caption{The Hall coefficient vs T for film samples of composition 
(La$_{1-y}$Gd$_y$)$_{0.67}$ Ca$_{0.33}$MnO$_3$, $y$ = 0 ($\square $) and $y$ = 0.25 
($\blacktriangle $). The dashed line correspond to a fit of the form 
$R_H=R_H^0\exp [E_H/k_BT]$. Inset a): $1/R_H$ vs $T$, for $y$ = 0.25
 showing an extrapolation to 245 K $>>T_C$ = 142 K. Inset b) $\ln |RH|$ vs 1000/T, 
for $y$ = 0.25, showing an activation energy $E_H=91\pm 5$ meV.}
\label{ fig10}
\end{figure}%

\begin{figure}
\caption{ Left: the resistivity vs $T^2$ for magnetic fields up to 70 kOe in a single 
crystal sample of composition La$_{0.67}$(Ca,Pb)$_{0.33}$MnO$_3$. 
Right: the magnetization deviation from saturation vs temperature on a log-log plot, 
with $T^{3/2}$ and $T^{5/2}$ contributions.}
\label{fig11 }
\end{figure}%

\begin{figure}
\caption{Left: the experimental resistivity $\rho (T)$ after substraction of the residulal 
value $\rho _0^{exp}$, divited by $\alpha _HT^2$vs temperature. The dashed line is 
the result of the one-magnon calculation described in the text. Right: $\rho (T)$ after 
substraction of the fitted value $\rho _0^{fit}$ vs temperature in a log-log plot. In both 
plots significant deviations from the $T^2$ behavior displayed in Fig. 12 are observed.}
\label{fig12 }
\end{figure}%

\begin{figure}
\caption{The Seebeck coeficient $S$ vs temperature. It is positive and metal-like at 
low temperature, has an anomalous kink near 30 K, and develops a positive field 
dependence above 40 K. The dashed line is a linear fit in the low temperature regime.}
\label{ fig13}
\end{figure}%

\begin{figure}
\caption{a) Resistivity vs temperature in zero field. The broken lines indicate 
extrapolations of the fits to the low and high temperature regions of the curve. 
The solid line is the parallel combination of the two conductivities using the 
magnetization (inset) as a mixing factor. b) Similar results for the Seebeck coefficient.}
\label{ fig14}
\end{figure}%

\begin{figure}
\caption{a) Magnetization data for this sample. b) Resistivity data as functions of field 
and temperature. The dashed curve is a parallel admixture using the reduced 
magnetization measured at 10 kOe.}
\label{fig15 }
\end{figure}%

\begin{figure}
\caption{a) The mixing factor $c(H,T)$ extracted from the resistivity as described in 
the text. b) Seebeck coefficient data and results of a computation using $c(H,T)$ from 
a) in Eq. 8.}
\label{fig16 }
\end{figure}%

\begin{figure}
\caption{a) the mixing factor $c(H,T)$ calculated in the mean-field model with 
$\alpha$ =0.02 and $\gamma$ =0.3. b) The magnetization calculated with the same 
parameters. The dotted line shows the non-interactive case for comparison.}
\label{ fig17}
\end{figure}%

\begin{figure}
\caption{a) Inverse magnetic susceptibility H/m vs temperature near the M-I transition. 
The appearance of free carriers induce the rise of the effective $T_C$, in qualitative 
agreement with data by Goodwing et al. b) Calculated resistivity using $c(H,T)$ from 
Fig. 17a) in Eq. 7.}
\label{ fig18}
\end{figure}%

\end{document}